\documentclass[prl,twocolumn,floatfix,amsmath,amssymb,showpacs,superscriptaddress]{revtex4}
\usepackage{graphicx,color}
\usepackage{dcolumn}
\usepackage{bm}

\begin{document}

\title{Nonuniversal routes to universality: Critical phenomena in colloidal 
dispersions}

\author{D.~Pini} 
\affiliation{Dipartimento di Fisica, Universit\`a di Milano, Via Celoria 16, 20133 Milano, Italy}
\author{F.~Lo~Verso}
\affiliation{Institut f\"ur Theoretische Physik II, Heinrich-Heine-Universit\"at
D\"usseldorf, Universit\"atsstra{\ss}e 1, D-40225 D\"usseldorf, Germany}
\altaffiliation[Current address: ]{Chimie Analytique et Biophysico-chimie de l'Environnement (CABE), Universite de Geneve, Sciences II,
30 quai E. Ansermet, CH-1211 Geneve 4, Switzerland.}
\author{M.~Tau}
\affiliation{Dipartimento di Fisica, Universit\`a di Parma, Parco Area delle Scienze 7/A, 43100 Parma, Italy}
\author{A.~Parola}
\affiliation{Dipartimento di Scienze Fisiche, Universit\`a dell'Insubria, Via Valleggio 11, 22100 Como, Italy}
\author{L.~Reatto}
\affiliation{Dipartimento di Fisica, Universit\`a di Milano, Via Celoria 16, 20133 Milano, Italy}

\begin{abstract}

We investigate critical phenomena in colloids by means 
of the renormalization-group based hierarchical reference theory 
of fluids (HRT). We focus on three experimentally relevant 
model systems, namely the Asakura-Oosawa model 
of a colloidal dispersion under the influence
of polymer-induced attractive depletion forces; fluids with competing short-range 
attractive and longer-range repulsive interactions; 
solutions of star-polymers whose pair potential presents both
an attractive well and an ultrasoft repulsion at shorter distance.
Our results show that the ability to tune the effective interactions 
between colloidal particles allows one to generate a variety 
of crossovers to the asymptotic critical behavior, which are 
not observed in atomic fluids. 

\end{abstract}

\maketitle
The description of colloidal dispersions in terms of effective interactions
acting between the macroparticles~\cite{belloni,likos} has opened the realm 
of complex fluids to
modern Liquid-State Physics, and is the main reason for the rekindled interest
in this discipline. On the one hand, using colloidal particles as macroatoms 
allows visual inspection of the fluid, solid and glassy states of matter which 
occur also in atomic substances~\cite{pusey}. On the other hand, the diversity 
of effective 
interactions, as well as the possibility of tuning them by changing the 
parameters which characterize the dispersion, leads to a phase behavior which 
is much richer than that encountered in atomic fluids~\cite{likos}. 
This ability of engineering the phase diagram~\cite{yethiraj} is fascinating 
both in its own sake, and for its potential in technology. 
Compared to the study of their overall phase behavior, critical phenomena 
in colloids have received less attention. 
A likely reason for this situation is that, 
in contrast with the menagerie 
of possible phase equilibria, the most relevant feature of critical phenomena
is their universality, as expressed by the critical indexes being the same 
for all the systems belonging to the same universality class. For fluid-fluid
(FF) transitions, this is expected to be the Ising one for both simple and 
colloidal fluids such as protein solutions
and colloid-polymer mixtures,
so it might be argued that there is little to learn in going 
from simple fluids to colloids. 
However, while universality comes along 
asymptotically close to the critical point, the approach to the asymptotic 
regime is by no means universal:  
how the asymptotic critical regime is reached, 
and how far from the critical point critical fluctuations are expected 
to become important, 
are relevant questions whose answers can be quite different from those
found in atomic fluids, since 
the specific features of colloid-colloid effective potentials may induce 
peculiar crossovers, and affect the size 
of the critical region. 

In this Letter, we show how the onset of critical fluctuations is affected
by the specific form of the interaction by focusing on three model systems,
all of which have received much attention because of their phase behavior 
and structural properties: the Asakura-Oosawa (AO) pair potential 
for hard-core colloidal 
particles in the presence of non-adsorbing polymer in solution; a hard-core
two-Yukawa (HCTY) fluid with
competing short-range attractive and longer-range repulsive interactions; 
a model of a star-polymer (SP)
solution, where the ultrasoft repulsion between
the stars is followed by an attractive well. 
We have investigated their
critical behavior by the hierarchical reference theory (HRT)~\cite{hrt}. 
In HRT, 
the attractive part of the interaction in Fourier space is gradually switched
on by introducing an evolving infrared cut-off $Q$ such that at any 
intermediate stage of the process, fluctuations with characteristic 
length $L\agt Q$ are strongly suppressed.  
The fully interacting system and the onset of long-range 
correlations are then recovered in the limit $Q\to 0$. 
HRT is optimally suited to the present study because it implements 
the renormalization group (RG) method, and therefore yields a realistic, 
non mean-field-like treatment of the long-wavelength fluctuations which drive 
criticality and phase separation. 
On the other hand, this approach 
keeps all the information about the non-universal properties which 
are expected to depend 
on the specific features of the interaction, and would be lost 
in a coarse-graining approach. The crossover to the asymptotic power-law
behavior can be described by introducing an ``effective'' critical
exponent, defined as the local slope of the quantity in hand as a function 
of the reduced temperature $t$ in a log-log plot. In the following, 
we will adopt the standard definition $t=(T-T_{c})/T_{c}$, $T_{c}$ being the
critical temperature.  

We first consider the AO model. The AO effective pair potential 
of polymer-induced depletion forces~\cite{ao} consists of an attractive tail
which vanishes identically beyond the polymer diameter, measured from the 
hard-core surface. The  
strength of the interaction is controlled  
by the fugacity $z_{p}$ 
of the polymer or, equivalently, by the packing fraction $\eta^{r}_{p}$ 
of the pure polymer solution in osmotic equilibrium with the colloid-polymer
mixture. This ``reservoir'' packing fraction plays the role of the inverse
temperature in a thermal fluid. 
Consistently with the above definition of $t$, we have set 
$t=(\eta^{r}_{p,{\rm crit}}-\eta^{r}_{p})/\eta^{r}_{p}$, 
$\eta^{r}_{p,{\rm crit}}$ being the critical value of $\eta^{r}_{p}$.  
A HRT study of the overall phase diagram
and asymptotic critical behavior of several thermodynamic 
quantities has been presented elsewhere~\cite{hrtao1,hrtao2}.
However, the approach to criticality has not been studied before.
\begin{figure}
\includegraphics[height=6.cm,angle=0.,clip]{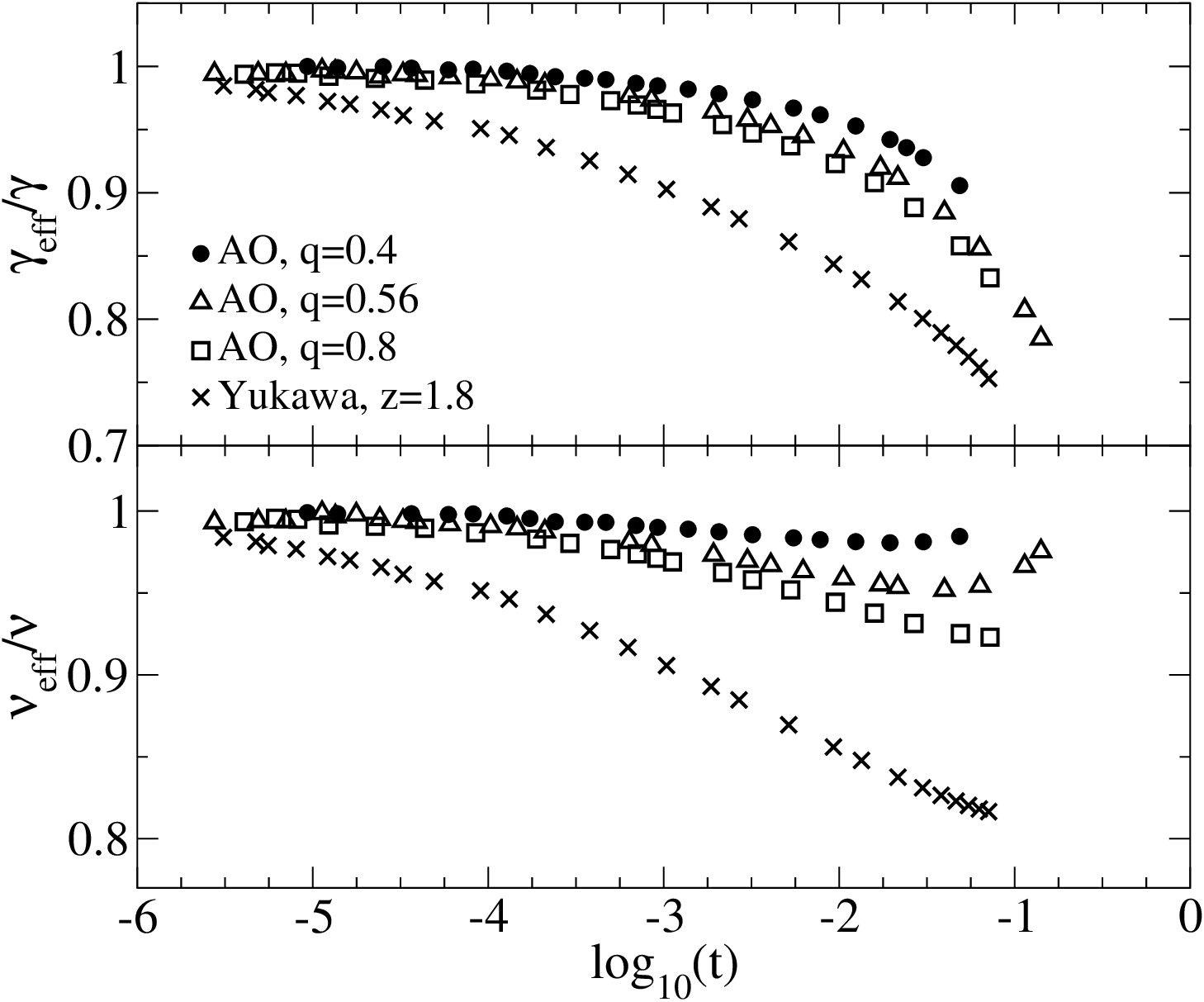}
\caption{ 
AO effective exponents $\gamma_{\rm eff}$ 
(upper panel) and $\nu_{\rm eff}$ (lower panel) for $\chi_{\rm red}$ 
and $\xi$,
normalized by their asymptotic values, compared with the effective exponents
of an attractive HCY potential with inverse-range parameter $z=1.8$.
For the AO potential, 
the reduced  effective temperature $t$ is 
$(\eta^{r}_{p,{\rm crit}}-\eta^{r}_{p})/\eta^{r}_{p}$, where
$\eta^{r}_{p}$ is the reservoir packing fraction. 
}
\label{fig:ao}
\end{figure}
In Fig.~\ref{fig:ao} we have plotted, for several polymer-to-colloid size ratios
$q$, the effective exponents 
$\gamma_{\rm eff}$, $\nu_{\rm eff}$ 
normalized by their asymptotic values $\gamma$, $\nu$, which describe 
the divergence of the (reduced) compressibility 
$\chi_{\rm red}$ and correlation length $\xi$ on the critical isochore as 
$T_{c}$ is reached. In HRT one has  
$\gamma=1.378$, $\nu=\gamma/2=0.689$. 
Figure~\ref{fig:ao} also shows the effective exponents
of a hard-core Yukawa (HCY) potential with attractive tail and 
inverse-range parameter $z=1.8$, which is 
appropriate to model the pair potential of simple atomic fluids 
(inverse ranges are in units of $\sigma^{-1}$, $\sigma$ 
being the hard-sphere diameter). 
As expected
on the basis of the RG predictions~\cite{fisher}, as $q$ is decreased, the asymptotic
critical regime is reached at larger reduced temperatures, and  
for $q=0.4$, $\gamma_{\rm eff}$ is about $95\%$ of its asymptotic value already 
at reduced effective temperatures $t\sim 0.1$. This is different from what is 
observed in the HCY fluid, whose $\gamma_{\rm eff}$ at these reduced 
temperatures is still mean-field-like. 
We did not consider size
ratios $q<0.4$, since our implementation of HRT is not suited to very narrow 
potentials. 
Moreover, in the AO potential, FF phase separation is preempted by freezing
for $q\lesssim 0.4$~\cite{dijkstra,hrtao1}, even though at smaller $q$ it can still 
be observed as a metastable transition. 
For the values of $q$ considered here, the AO depletion interaction has 
many-body contributions, which are not accounted for by the effective pair 
potential which we have used. However, the critical behavior is captured 
satisfactorily already at the pair potential level~\cite{hrtao2}. 
On the basis of our results, we expect AO colloid-polymer mixtures to exhibit 
nearly asymptotic Ising scaling at reduced effective temperatures 
$t\simeq 0.1$. 
We also note that there is no exponent renormalization due to the binary 
nature of the system,
provided the {\em reservoir} packing 
fraction is used as the relevant field driving criticality. 

In a recent paper~\cite{royall}, the critical behavior of the correlation 
length of colloid-polymer mixtures has been 
investigated experimentally, 
and the critical exponent $\nu$ was found to be close to the mean-field value
$\nu=0.5$. 
On the other hand, Monte Carlo studies supplemented by finite-size scaling 
unambiguously proved that the AO model belongs to the Ising universality
class~\cite{hrtao2,horbach}. This rises~\cite{reatto} an interesting question 
about the possible presence of an unforeseen long-range interaction 
in the mixtures studied in Ref.~\cite{royall}. 

Besides the tunable short-range attractions of which the AO potential represents the best-known example, fluids where the short-range attraction is followed by a repulsion which 
takes over at long distance have been receiving increasing attention. 
It has long been 
acknowledged~\cite{brazovskii} that the competition between attraction and repulsion leads, for strong 
enough repulsion, to the disappearance of bulk 
FF phase separation. 
As the temperature is lowered, this is replaced by the occurrence 
of nonhomogeneous phases that may have different morphologies~\cite{seul},
e.g. cluster- or stripe-shaped.
\begin{figure}
\includegraphics[height=7.cm,angle=0.,clip]{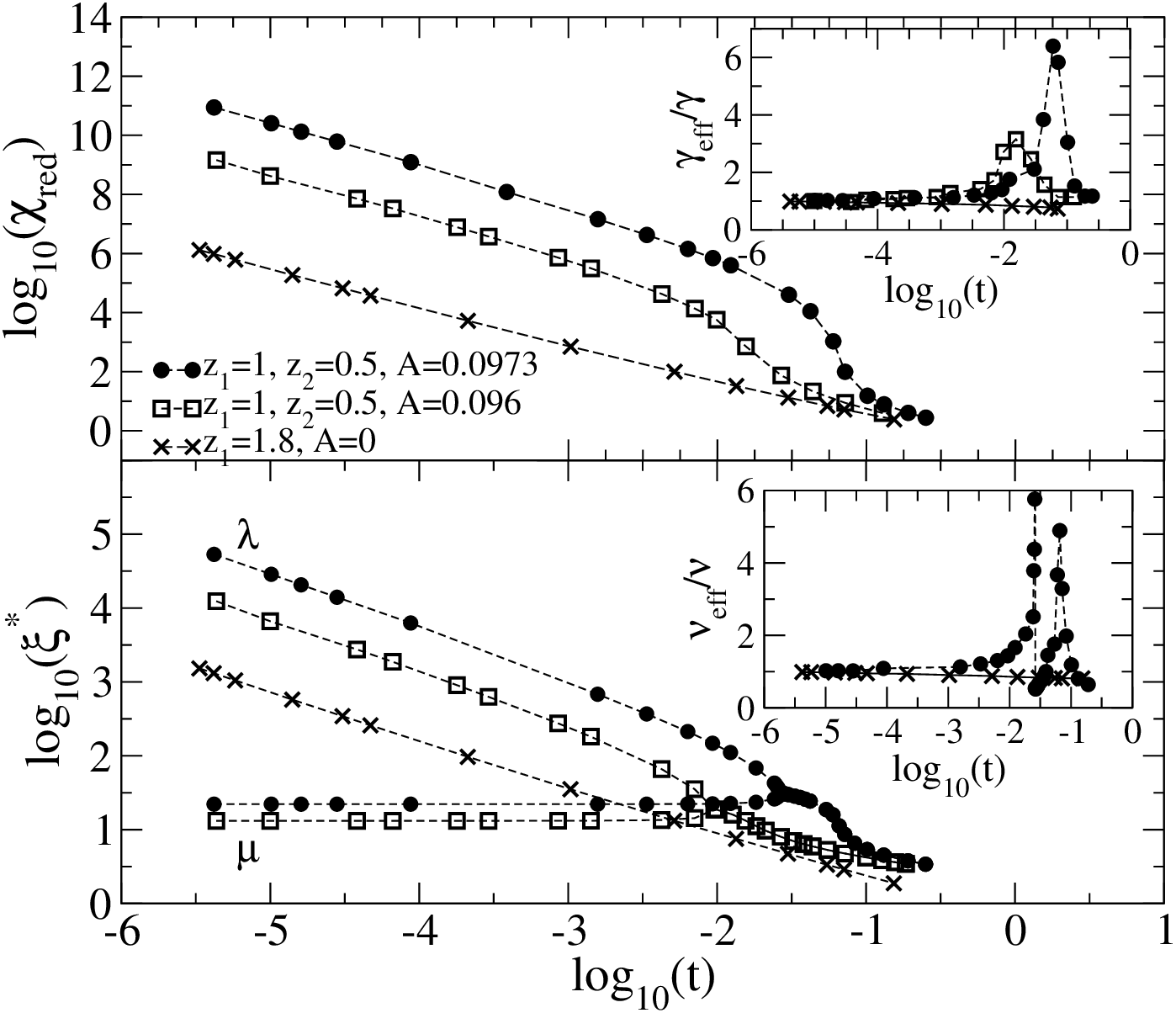}
\caption{ 
Reduced compressibility (upper panel) 
and correlation 
length (lower panel) in units of the particle diameter 
of a HCTY fluid 
with competing short-range attraction
and longer-range repulsion.
The insets show 
$\gamma_{\rm eff}$, 
$\nu_{\rm eff}$ compared with the effective exponents
of a HCY potential 
($\nu_{\rm eff}$ for $A=0.096$ has been omitted for clarity).
See text for the meaning of the lengths $\lambda$ and $\mu$. 
In the oscillatory regime, only the real part of $\lambda$ and $\mu$ 
has been shown. Lines are a guide for the eye.}
\label{fig:compe}
\end{figure}

Instead of considering the nonhomogeneous phases, here we focus 
on a regime of relatively weak
repulsion, such that the competition is not so strong as to cause the disappearance 
of FF phase separation, and investigate the effect on the critical behavior. 
Our model interaction is the HCTY potential 
with inverse-range parameters
$z_{1}=1$, $z_{2}=0.5$  for the attractive 
and repulsive contribution respectively~\cite{chemphyslet}. 
The relative amplitudes $A$ of the repulsion is close
to the value above which the FF transition disappears, 
and microphase formation
is expected. This regime is characterized by an extremely flat FF coexistence
curve~\cite{chemphyslet}.
In Fig.~\ref{fig:compe},  
$\chi_{\rm red}$ and $\xi$ have been plotted together with 
the effective critical exponents. 
The most striking effect is the steep increase of $\chi_{\rm red}$ and $\xi$ 
at reduced temperatures $t\simeq 0.1$, with a corresponding sharp crossover 
of the effective exponents. 
The enhancement of large density fluctuations 
can be traced back to the tendency towards cluster formation triggered 
by the competition.
A sign of this ``incipient clustering'' is that, unlike in Lennard-Jones 
(LJ)-like fluids 
with purely attractive tail potentials, one finds damped oscillatory decay 
for the correlations even for states very close to critical~\cite{archer}. 
On further approaching 
the critical point, the decay becomes monotonic, but takes a double exponential form
described by {\em two} characteristic lengths $\lambda$, 
$\mu$~\cite{chemphyslet,freezcomp}. 
A similar behavior has in fact been found in many frustrated 
systems~\cite{nussinov}. In our treatment, $\lambda$ and $\mu$ 
are obtained from expanding the reciprocal of the structure
factor $S(k)$ at small wave vector $k$ up to $k^{4}$.
Asymptotically close
to the critical temperature, $\lambda$ reduces to the usual Ornstein-Zernike (OZ) 
correlation
length, while $\mu$ saturates at a finite value. The change from oscillatory to monotonic
behavior is marked by the bifurcation in Fig.~\ref{fig:compe}, where $\lambda$, $\mu$ 
turn from complex conjugate to real. This is an example of 
the well-known 
Kirkwood crossover~\cite{kirkwood}. 
We observe that $\nu_{\rm eff}$ shows both a peak at a temperature slightly 
higher than that of the bifurcation, and a divergence  
at the low-temperature side of the bifurcation, where $\lambda$ takes 
a vertical slope. 
The latter feature is due to the coalescence of $\lambda$ and $\mu$ 
at the Kirkwood line.

It is tempting to intepret the above scenario as a cluster-driven
criticality, whereby particles are first strongly correlated within regions of size 
$\sim \mu$, which in turn become correlated over length scales of size $\sim \lambda$ 
as the critical point is reached. In the asymptotic critical regime, $\chi_{\rm red}$
diverges as $\lambda^{2}$ like in the OZ picture, but with an amplitude 
$\sim \mu^{2}$, much bigger than that of LJ-like fluids, 
see Fig.~\ref{fig:compe}. The possibility of clusters behaving like super-atoms
has already been pointed out in the context of the glass 
transition~\cite{zaccarelli}. 
In our case, however, clusters denote regions of enhanced particle-particle correlations,
rather than stable particle aggregates. 

The anomalous crossover of the HCTY model close to the stability limit of 
the liquid-vapor transition has not been observed experimentally yet. However,
it seems likely that such a regime can be experimentally reached. 
As an example, let us consider the system studied 
by Cardinaux {\it et al.}~\cite{cardinaux} 
as a model of lysozyme solutions.  
They investigated cluster formation for volume fractions $\eta$ 
of lysozyme in the range $0.085\leq \eta\leq 0.201$ in a solution of sodium 
hydroxyde at concentration $c=8{\rm mM}$. An estimate of the electrolyte 
concentration below which the liquid-liquid transition is expected to disappear can be 
obtained within the random phase approximation by locating the concentration 
at which the convexity 
of the Fourier transform of the potential at zero wave vector $k$ turns from positive 
to negative. 
According to this criterion, 
$c=8{\rm mM}$ is indeed well inside the cluster region, while at $\eta=0.201$, 
the threshold concentration is at $c\simeq 56 {\rm mM}$. 

\begin{figure}
\includegraphics[height=5.5cm,angle=0.,clip]{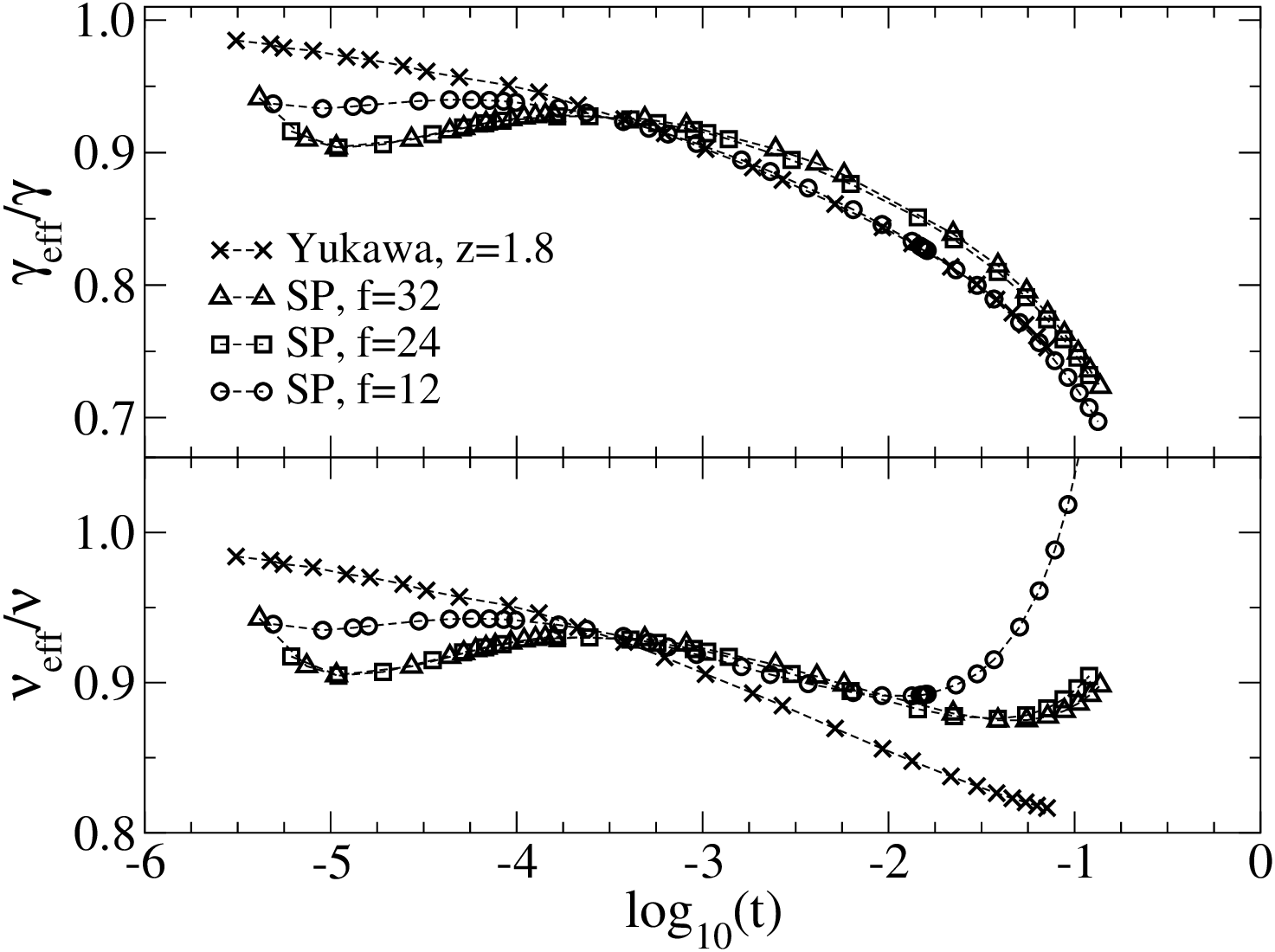}
\caption{ 
Same as Fig.~\protect\ref{fig:ao} for 
three SP solutions. Lines are a guide for the eye.}
\label{fig:star}
\end{figure}

Star-polymer solutions differ from the two systems considered above inasmuch 
as the interaction which accounts for the excluded volume effect is not the hard-sphere 
potential, but an ultrasoft repulsion of entropic origin with a logarithmic core and 
a repulsive Yukawa tail~\cite{likos2}. 
The size of the star $\sigma$ 
is of the order of the radius of gyration.
It has been shown that, when an additional 
attractive interaction between the stars is present, the solution may display two 
FF transitions, each ending at a critical point~\cite{staratt}. 
Here we focus on the behavior 
in the neighborhood of the lower-density critical point, which we expect to be more easily 
accessible by experiments. 
We considered three values of the arm number or ``functionality''
$f$, namely $f\! =\! 12$, $f\! =\! 24$, $f\! =\! 32$.
The attractive well was modeled by the same Fermi function used in Ref.~\cite{staratt}. 
Figure~\ref{fig:star} shows the effective exponents $\gamma_{\rm eff}$, $\nu_{\rm eff}$. 
The main qualitative difference with respect to the simple-fluid case represented 
by the HCY fluid with $z=1.8$ consists in the non-monotonic behavior 
of the effective exponents. As one moves away from the critical temperature, 
$\nu_{\rm eff}$ shows a minimum in the interval $10^{-2}<t<10^{-1}$, 
and increases sharply as $t$ is further increased. 
This is due to the effect of the soft repulsion which, at the small densities
considered here, is more substantial than in a hard-core system of comparable
size. 
Moreover, both $\gamma_{\rm eff}$ and 
$\nu_{\rm eff}$ show another minimum at $t\simeq 10^{-5}$ and bend towards 
their 
asymptotic values as $t$ is further reduced. The accuracy limit entailed 
by our numerical calculation prevented us from reaching reduced temperatures smaller than
$t\sim 10^{-6}$, so that the asymptotic exponents could not be observed. 
Notwithstanding this, the quantitative deviations from simple-fluid behavior 
for $t\lesssim 10^{-2}$
are rather 
small, and the effective exponents remain well apart from their mean-field values even
in the region of the minimum, at least for the values of $f$ investigated here. 
This behavior is different 
from what is found in the linear-polymer solutions considered 
in Ref.~\cite{bates}. In that case, the correlation length $\xi$ away from 
the critical point is considerably smaller than the polymer size $\sigma$. 
As the critical point is approached and $\xi$ becomes comparable to $\sigma$, 
the solution shows a rather sharp crossover from mean-field to Ising critical behavior. 
Here, on the other hand, 
the size of the star $\sigma$ is the natural lengthscale of both 
the repulsive and attractive interaction, so that $\xi$
is of the order $\sigma$ even away from the critical point.   
This requires that
the quality of the solvent should be unchanged at the transition, so that
the inter-star attraction is not overwhelmed by the attraction
between monomers within the same star.
Experimentally this condition can be reproduced by
considering effective depletion attractions between stars  
in multicomponent star-chain~\cite{stiakakis} or star-star
mixtures~\cite{mayer}.
These systems have been looked upon
for a big component in solution
with higher $f$ than those we investigated here.
Nevertheless, from the trends evidenced and
the comparison between the effective interaction shapes,
we expect
the appearence of a stable low-density FF transition
for highly or moderately asymmetric mixtures with small $f$ 
($f\ll 50$) for the big component.

In summary, we have presented an investigation of the critical behavior 
of three model fluids, 
whose pair interactions are relevant for a wide class of complex liquids,
although they certainly do not aim at covering the whole subject of criticality 
in colloidal systems. In the AO model,
nearly asymptotic Ising-like behavior is observed in a wide interval of reduced 
effective temperatures for the polymer-colloid size ratios at which 
the effective AO pair 
interaction is a reliable representation of the binary colloid-polymer mixture. 
In fluids with competing interactions, the enhancement of density fluctuations due 
to competition causes strong crossovers with anomalously high effective critical exponents
$\gamma_{\rm eff}$, $\nu_{\rm eff}$ and critical amplitudes, as well as the emergence of two 
characteristic lengths. In star-polymer solutions with star-star attraction, we find 
a non-monotonic behavior of the effective exponents $\gamma_{\rm eff}$, $\nu_{\rm eff}$ 
close to the critical temperature. 
We expect that these predictions can be tested by experiments  
in colloid-polymer and polymer-polymer mixtures and protein solutions.  

This work is funded in part by the Marie Curie program of the
European Union, contract number MRTN-CT2003-504712, and the Foundation 
BLANCEFLOR Boncompagni-Ludovisi, n\'ee Bildt. FLV gratefully 
acknowledges discussion with Christos Likos and Christian Mayer.

\end{document}